\documentclass[%
,twocolumn%
,amsmath, nobibnotes, aps, pre]{revtex4-1}
\usepackage{bm}%
\expandafter\ifx\csname package@font\endcsname\relax\else
 \expandafter\expandafter
 \expandafter\usepackage
 \expandafter\expandafter
 \expandafter{\csname package@font\endcsname}%
\fi

\usepackage{graphicx,subfigure}
\usepackage{amssymb,amsmath,wasysym}

\DeclareGraphicsExtensions{.pdf,.png}
\graphicspath{{./figs_pre/}{./}}
\usepackage[usenames,dvipsnames]{color}   

\usepackage[normalem]{ulem} 

\def\be{\begin{equation}}
\def\ee{\end{equation}}
\def\bpm{\begin{pmatrix}}
\def\epm{\end{pmatrix}}

\renewcommand{\P}{\mathcal{P}}
\newcommand{\K}{\mathcal{K}}

\newcommand{\FE}{\mathcal{F}}
\newcommand{\V}{\Omega}
\newcommand{\A}{A}
\newcommand{\h}{H}
\newcommand{\w}{W}

\renewcommand{\L}{\mathcal{L}}

\begin{document}

\title{Mechanical stress as a regulator of cell motility}

\author{T. Putelat$^{1}$, P. Recho$^{2}$ and L. Truskinovsky$^3$}
\affiliation{
$^1$DEM, Queen's School of Engineering, University of Bristol, 
Bristol, BS8 1TR, United Kingdom\\
$^2$ LIPhy, CNRS--UMR 5588, Universit\'e Grenoble Alpes, F-38000 Grenoble, France\\
$^3$PMMH, CNRS--UMR 7636, ESPCI PSL, F-75005 Paris, France
}
\email{\mbox{t.putelat@bristol.ac.uk, pierre.recho@univ-grenoble-alpes.fr,} lev.truskinovsky@espci.fr}

\date{\today}%

\begin{abstract} \small  
The motility of a cell can be triggered or inhibited not only by an applied force but also by a mechanically neutral force couple. This type of loading,  represented by an applied stress and commonly interpreted as either squeezing or stretching, can originate from extrinsic interaction of a cell with its neighbors. To quantify the effect of applied stresses on cell motility we use an analytically transparent one-dimensional model  accounting for active myosin contraction and induced actin turnover. We show that stretching can polarize static cells and initiate cell motility while squeezing can symmetrize and arrest moving cells. We show  further that sufficiently strong squeezing can lead to the loss of cell integrity. The overall behavior of the system depends on the two dimensionless parameters characterizing internal driving (chemical activity) and external loading (applied stress). We construct a phase diagram in this parameter space distinguishing between, static, motile and collapsed states. The obtained results are   relevant for  the mechanical understanding of contact inhibition and the epithelial-to-mesenchymal transition. 
\end{abstract}


\maketitle

\section{Introduction}

Cell migration plays a key role in ensuring the development, integrity and regeneration of living organisms \cite{alberts2014molecular}. Fueled by ATP hydrolysis, cells can self-propel in specific directions due to intricate biochemical and genetic regulation. Cell motility can also be controlled by resultant mechanical forces as it was established in experiments addressing motility initiation and motility arrest \cite{Verkhovsky1999,anderson1996coordination, Heinemann2011, Schreiber2010} and is exemplified by force-velocity relations~\cite{RecTru_pre13}. 

In this paper we study  another mechanical regulation mechanism through balanced force couples that can either \emph{squeeze} or \emph{stretch} a cell. The importance of such loading conditions, represented by an applied stress, is corroborated by the fact that cells mostly exist in crowded and therefore  mechanically constrained  environments and, that essential physiological functions, such as wound healing and tissue regeneration, take place due to collective cell migration~\cite{FriDar_nrm09, hakim2017collective, camley2017physical}. 

There exists considerable experimental evidence from guided migration of cell monolayers on a substrate~\cite{FriDar_nrm09,Trep_tcb11,Gov_nmat11} indicating the presence of a mechanical feedback mediated not only by the pulling forces exerted by leader cells, and traction forces from the substrates, but also by the transmission of mechanical stress through intercellular junctions~\cite{Tre_nphys09,Gov_hfspj09,Kab_jrsi12,Tam_nmat11,deforet2014emergence, garcia2015physics}.  For instance, stresses appear to be responsible for the fact that cells in confined proliferating monolayers cease their motility when they reach confluence \cite{garcia2015physics}, a phenomenon known as contact inhibition~(CI)~\cite{stramer2016mechanisms}. Stresses are also involved in the epithelial-to-mesenchymal transition (EMT) when destabilization of epithelial layers through the loosening of cell-cell contacts results in an increased cell mobility and ultimately leads to invasive and metastatic behavior~\cite{gjorevski2012regulation}. 

Several experimental protocols, allowing one to stretch or squeeze a cell by externally applied force couples, are currently available, including optical tweezers \cite{zhang2008optical}, microfluidic devices \cite{gossett2012hydrodynamic}, atomic force microscopy \cite{watanabe2011direct} and photothermally activated micropillars \cite{sutton2017photothermally}. Experiments involving these techniques confirm that stretching is not only an important determinant of the motility status but also a potential regulator of cell differentiation or death~\cite{lee1999regulation, vogel2006local, ao2015stretching, gudipaty2017mechanical}.  
A typical explanation of such observations  relies on mechanics only indirectly. For instance, the mechanosensitive nature of ion channels~\cite{lee1999regulation, tao2015active, Taloni2015} is used as a justification that stretching affects fluxes across the cell membrane. The latter can be responsible for an increased expression of small RhoGTPase (Rho, Rac, Cdc42)  regulating  the behavior of the cytoskeleton \cite{goldyn2009force, franze2009neurite}. In the case of EMT, activation of Rho is expected to provoke the nuclear translocation of transcription factors, which promote the expression of EMT-regulating genes controlling the disassembly of cell-cell contacts \cite{morita2007dual}. 

In this paper we show that a more direct mechanical interpretation of some of these experimental observations can be obtained from the study of a one-dimensional model of an externally stressed  cell crawling on a rigid substrate. A prototypical example of this motility mechanism is provided by cells self-propelling inside rigid channels \cite{DoyWanMatYam_jcb09,maiuri2012first}. The functioning of the mechanical machinery involved in cells crawling is rather well understood \cite{JulKruProJoa_pr07, Rubinstein2009, mogilner2009mathematics, shao2010computational, doubrovinski2011cell, ziebert2011model, RecPutTru_prl13, giomi2014spontaneous, tjhung2015minimal}. In particular the question of how such cells sense gradients and direct their motion over large distances has also been thoroughly studied \cite{Banerjee2011, leberre2013geometric, comelles2014cells, prentice2016directional}. However, the role of an applied stress still needs to be elucidated. 

To highlight the role of stresses in an analytically transparent setting, we represent the cell as an active segment limited by elastically interacting moving boundaries \cite{JulKruProJoa_pr07,Rubinstein2009,Barnhart2010,RecPutTru_prl13}. We develop a version of the active gel theory where actin density is controlled homeostatically and assume that the internal flow generation, implying actin turnover, is driven exclusively by myosin contraction \cite{heisenberg2013forces,RecPutTru_jmps15}; this description is particularly relevant for bulk cells in a tissue which can only produce limited protrusions \cite{hakim2017collective}. We study in this setting the effect on motility of an externally applied mechanical couple with \emph{zero} resultant. The analysis of the role of the resultant can be found in a companion paper \cite{CIL_paper}.

Our main finding is that mechanical stretching can polarize static cells and initiate their motility while squeezing can symmetrize and arrest moving cells. Depending on the amount of the applied stress, the system exhibits three  states: collapsed (cell death or division), static (symmetric and passive) and motile (polarized and active). The peculiar feature of the ensuing phase diagram, with one axis representing contraction and another, characterizing applied stress, is the fact that the transition between static and collapsed states is discontinuous while the transition between static and motile states is continuous. Interestingly, the critical end point separating the first order transitions from the second order transitions is located in a physiologically relevant part of the diagram. 

Our general conclusion is that motility is favored by strong contraction and weak squeezing  while sufficiently strong squeezing leads to collapse independently of the strength of the contraction. In the competition between passive tension and active contraction, symmetric immobile configurations represent a delicate balance. Another conceptual result  is that the effect of the homeostatic regulation of actin density on motility initiation and cell collapse is rather similar to the effect of the applied stress. The obtained quantitative relations between dimensionless parameters, characterizing various stability thresholds in this problem, may be relevant  for a broad range of biological phenomena including EMT and CI. 

The paper is organized as follows. In Sec.~\ref{s:modelintro} we formulate the model and identify 
three nondimensional parameters, 
which fully determine the behavior of the system. In Sec.~\ref{s:steady} we characterize the three distinct steady regimes describing static, collapsed and motile configurations. In Sec.~\ref{s:numstability} we present the phase or regime diagram in the space of dimensionless parameters and delineate the thresholds between different types of behavior. The nature of the implied transitions is elucidated in Sec.~\ref{s:discussion} where we also discuss the relevance of the obtained results for biological systems.  Section~\ref{s:conclusion} summarizes our findings and addresses some open problems. 
In Appendix A we introduce a natural extension of the model which regularizes the phenomenon of contractility-induced collapse. 
In Appendix B we develop analytical asymptotics for myosin distribution.

\section{The model}\label{s:modelintro}

\begin{figure}
\centering
\includegraphics[scale=0.4]{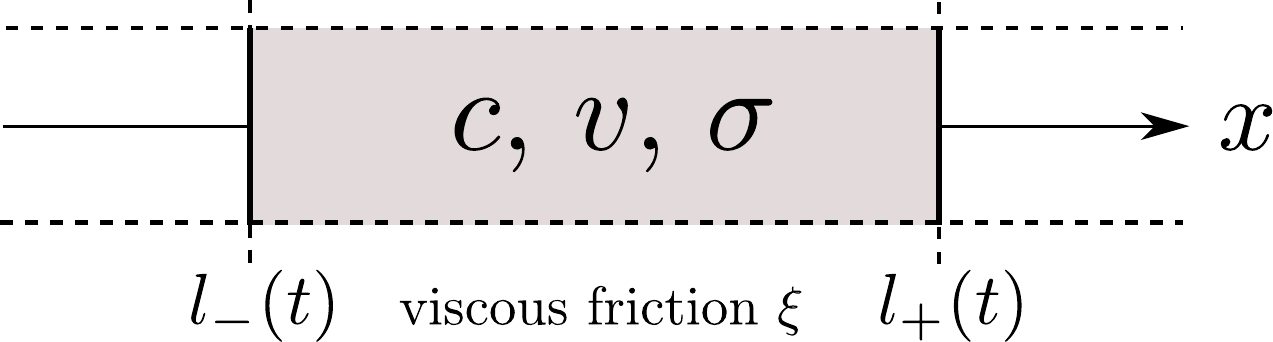}
\caption{Schematic representation of a segment of active gel moving on a track.}
\label{f:cell_on_track}
\end{figure}

Consider a prototypical one-dimensional model of a cell fragment confined to a thin channel or a track (see Fig.~\ref{f:cell_on_track}). 
It can be represented as a continuum segment \mbox{$x\in [l_-(t),l_+(t)]$} with two moving boundaries $l_-(t)$ and~$l_+(t)$.

\emph{Actin dynamics.} Slow, overdamped motion of the continuously repolymerizing actin is described by the mass and momentum balance equations
\begin{subequations}
\begin{align}
\rho\,\partial_x v &= (\rho_0-\rho)/\tau , \label{e:eq_stress2_a}\\
\partial_x \sigma &= \xi v, \label{e:eq_stress2_b}
\end{align}
\end{subequations}
where $v(x,t)$ is the velocity, $\sigma(x,t)$ is the stress and~$\rho(x,t)$ is the density of the filamentous \mbox{F-actin} meshwork. We denoted by $\xi$ the coefficient of viscous friction with the rigid environment and by $\tau$ the turnover time of \mbox{F-actin}. The homeostatic density at which the polymerization and depolymerization of \mbox{F-actin} are balanced is denoted $\rho_0$ \cite{recho2016growth}. We implicitly assumed that inertia is negligible compared with viscous friction ($\rho v^{-1}dv/dt\ll\xi $) and that the density variation of \mbox{F-actin} in a material particle is small compared to the rate of its chemical turnover ($\rho^{-1}d\rho/dt\ll\tau^{-1}$).  A different model was considered in \citep{RecPutTru_prl13,RecPutTru_jmps15} where the total mass of \mbox{F-actin} was controlled instead of the target density. In the typical case when actin turnover is distributed inside the cell, rather than being narrowly localized on the boundaries, the present model appears to be more realistic.

Assume next that the internal stress can be represented as a sum of two terms 
\be
\sigma(\rho,c)=\sigma_a (\rho) +\sigma_m (c),
\label{e:eq_stress0}
\ee
where $\sigma_a $ is a contribution due to the compressibility of \mbox{F-actin} meshwork and $\sigma_m$ is a contractile stress due to the presence of myosin~II. The myosin motors with concentration $c(x,t)$ actively cross-link F-actin filaments and follow their own dynamics which is detailed below.
 
To define $\sigma_a$, we further suppose that the actin density is close to its homeostatic value $\rho\sim\rho_0$ and therefore we can use  the linear approximation $\sigma_a (\rho) \sim -p(\rho_0)-\partial_\rho p(\rho_0)  (\rho-\rho_0)$, where $p(\rho_0 )$ is the pressure in the homeostatic \mbox{F-actin} meshwork. At the same level of approximation, the mass balance equation \eqref{e:eq_stress2_a} reads $\tau\rho_0 \partial_xv \sim \rho-\rho_0$. Combining these two results we obtain the approximate constitutive relation
\be
\label{e:eq_stress1}
\sigma_a   = -p_h+\eta \partial_x v,
\ee
where $p_h=p(\rho_0)$ is the homeostatic pressure and $\eta=\tau\rho_0 \partial_\rho p(\rho_0)$ is the effective bulk viscosity \cite{ranft2010fluidization}. Note that the viscosity may also have a different origin \cite{JulKruProJoa_pr07,RecTru_pre13}. 

\emph{Myosin dynamics.} For the stress generated by myosin contraction we assume that $\sigma_m=\chi c$, 
where $\chi>0$ is the constant contractility coefficient (see \citep{RecPutTru_jmps15} for 
a nonlinear extension with contractility saturation). 
To specify the dynamics of myosin we assume that the motors may be either unbound with concentration $n(x,t)$ or bound (in a stall state) to \mbox{F-actin} filaments with concentration~$c(x,t)$. Suppose for simplicity, that the myosin-actin attachment-detachment dynamics is modeled as a linear reaction with an attachment rate~$k_b$ and a detachment rate~$k_u$:
\begin{subequations}\label{e:AD_motors}
\begin{align}
\partial_t c+\partial_x(c v-D_c\partial_x c)  &= k_bn-k_uc , \label{e:AD_motors_a}\\ 
\partial_t n+\partial_x(n v-D_n \partial_x n) &= k_uc-k_bn . \label{e:AD_motors_b}
\end{align}
\end{subequations}
Here, we assumed that in addition to being advected by the flow of \mbox{F-actin}, the bound and unbound myosin motors may diffuse with diffusivities~$D_c$ and~$D_n$ respectively. 
Suppose that the attachment-detachment reaction is close to equilibrium such that $ n/c\sim  K_{bu}$ while $K_{bu}=k_u/k_b\ll 1$. Then, we can define $D:=K_{bu}D_n+D_c$ and combine Eqs.~\eqref{e:AD_motors} to obtain the effective advection-diffusion equation of the bound myosin motors
\be
\partial_t c+\partial_x(c v)=  D \partial_x^2 c.
\label{e:motors_conservation}
\ee
Equations~\eqref{e:eq_stress2_b} and~\eqref{e:motors_conservation}, supplemented by the active gel constitutive relation $\sigma=-p_h+\eta \partial_x v + \chi c$, form the closed system for the three unknown fields $v$, $\sigma$ and~$c$.

\emph{Boundary conditions.} 
As long as there is no F-actin flow through the cell boundaries, the protrusive activity of myosin is neglected, 
and the two kinematic conditions,
\be
\label{e:kinematic_bc}
\dot{l}_\pm=v(l_\pm(t),t) ,
\ee
determine the dynamics of the cell fronts; the dot denotes time derivative. Note that conditions \eqref{e:kinematic_bc} do not guarantee conservation of the total mass of actin because of the presence of a bulk exchange with the homeostatic reservoir. In contrast, similar no-flux conditions for myosin motors,
\be
\partial_x c(l_\pm(t),t) = 0 ,
\label{e:motors_conservation_bc}
\ee
ensure that the total amount of motors $M=\int_{l_-(t)}^{l_+(t)}c(x,t)dx$ is conserved. 
 
The regulation of cell motility by mechanical stress, which is the main subject of this paper,  
is implemented through the boundary conditions  
\be
\label{e:sgm_lpm}
\sigma(l_\pm(t),t) =  \tau_\text{i}+ \tau_\text{e} ,
\ee
where we separate internally  and externally  generated tractions. 

The first term $\tau_\text{i}$ describes the mechanism controlling the cell length through the membrane-cortex tension \cite{ambrosi2016mechanics, keren2011cell}. We assume for simplicity that  $\tau_\text{i}(t)=-kL(t)$ where  $L(t)=l_+(t)-l_-(t)$ is the cell length and $k$ is the stiffness~\cite{diz2013use}. 
In our one-dimensional description, cell length variations can be related to mechanically induced cell volume variations or changes of the cell shape. Both happen, for instance, during cell spreading \cite{GuoPegWei_pnas17}. The associated timescale is comparable to the one involved in cell motility (minutes to hours) \cite{JilEde_plos11}.

The second term $\tau_\text{e}$ has its origin outside the cell and is interpreted as \emph{stretching} if $\tau_\text{e}>0$ and as \emph{squeezing} if $\tau_\text{e}<0$. 
It can describe, for instance, the cadherin-mediated interactions of the cell either with its neighbors or with the extracellular environment~\cite{RecTru_pre13} and is one of the two main parameters of the problem.

Using Eq.~\eqref{e:sgm_lpm} and the cell constitutive law [Eqs.~\eqref{e:eq_stress0} and \eqref{e:eq_stress1}], we can rewrite the mechanical boundary conditions in the form
\be
\label{e:bc_stress}
\eta \partial_x v(l_\pm(t),t) + \chi c(l_\pm(t),t) = -k(L(t)-L_0).
\ee
Relations similar to Eq.~\eqref{e:bc_stress} have been previously introduced on phenomenological grounds in~\cite{shao2010computational,Barnhart2010,Du2012,Loosley2012,RecTru_pre13,lober2015collisions}. Here we move a bit further and specify the expression for the \emph{homeostatic} length
\be
L_0=p_h/k+\tau_{\text{e}}/k ,
\label{e:L0}
\ee
separating contributions due to internal and external regulation.
More specifically, the first term on the right-hand side of Eq.~\eqref{e:L0} represents the internal regulation of the cell length through passive turnover of actin. The second term accounts for external mechanical squeezing or stretching: through this term  the environment  can also affect the conditions of homeostasis. 

\emph{Nondimensionalisation.} Choosing $l_0:=\sqrt{\eta/\xi}$ as the characteristic scale of length, $t_0:=\eta/(\xi D)$ as the scale of time, $\sigma_0:=\xi D$ as the scale of stress and $c_0:= M/(\sqrt{\eta/\xi})$ as the scale of motor concentration, we can reformulate the system of governing equations and the boundary conditions in a nondimensional form. The dimensionless problem depends only on three parameters: 
\be
\mathcal{P}=\chi c_0/\sigma_0,
\label{e:Pcal}
\ee
characterizing the strength of myosin contractility, 
\be
\K=k/(\sigma_0 l_0),
\label{e:Kcal}
\ee
representing the stiffness of the cell's boundary and 
\be
\L= L_0/l_0,
\label{e:Lcal}
\ee
the ratio of the homeostatic length~$L_0$ to the hydrodynamic length ~$l_0$. 
Typical physiological values for the material and dimensionless parameters in this model are collected in Table~\ref{t:valpar}.

In this paper, our main focus is on the parameter $\L$, defined in Eq.~\eqref{e:Lcal}, which contains two contributions, one due to internal remodeling of the cytoskeleton, $p_h/(kl_0)$, and another due to the external mechanical action of the cell environment, $\tau_{\text{e}}/(kl_0)$. Our goal is to show that the parameter $\L$, corresponding to a scaled stress actually, plays a crucial role in regulating both the initiation and the inhibition of cell motility.

\begin{table}
\scriptsize
\begin{tabular}{lll}
\hline\hline
Name & Symbol & Typical value \\ 
\hline
Viscosity & $\eta$ & $10^5$ Pa s \cite{JulKruProJoa_pr07,Rubinstein2009}\\
Contractility & $\chi c_0$ & $10^3$ Pa \cite{JulKruProJoa_pr07,Rubinstein2009} \\
Stiffness & $k$  & $5\times 10^8$ $\text{Pa\,m}^{-1}$ \cite{RecTru_pre13}\\
Motors diffusion coefficient & $D$ & $10^{-13}$ $\text{m}^{2}\text{s}^{-1}$ \cite{RecPutTru_jmps15}\\
Viscous friction coefficient & $\xi$ & $10^{15}$ Pa~s~$\text{m}^{-2}$ \cite{[{To obtain this value of friction coefficient, the number recommended in~\cite{JulKruProJoa_pr07} was divided by the thickness of the lamellipode $h\sim1\mu$m. In~\cite{Barnhart2011}, the coefficient $\xi$ is defined in the same way as in the present paper; however, friction is assumed to be strongly dependent on the physical properties of the substrate, and our choice corresponds to the upper limit of the interval recommended in~\cite{Barnhart2011}}] ref_fric_coeff} \\ 
Homeostatic length & $L_0$ & $2\times 10^{-5}$ m \cite{Rubinstein2009, Barnhart2010} \\
\hline
Characteristic length & $l_0=\sqrt{\eta/\xi}$ & $10^{-5}$ m  \\
Characteristic time & $t_0=\eta/(\xi D)$ & $10^3$ s \\
Characteristic velocity & $v_0=L_0/t_0$ & $72$ $\mu \text{m}\,\text{h}^{-1}$ \\
Characteristic stress & $\sigma_0=\xi D$ & $10^2$ Pa \\
\hline
Contractility parameter & $\mathcal{P}=\chi c_0/\sigma_0$ & $10$ \\
Stiffness parameter & $\mathcal{K}=kl_0/\sigma_0$ & $100$ \\
Cell length parameter & $\L=L_0/l_0$ & $2$\\
\hline\hline
\end{tabular}
\caption{\small Estimates of material coefficients and nondimensional parameter definitions.\label{t:valpar}}
\end{table}

\emph{Dimensionless system.} For simplicity, we do not relabel the dimensionless variables and we map 
the free boundary problem into a time-independent domain by introducing the comoving coordinate $y=[x-l_-(t)]/L(t)$. 
The main system of equations takes the form
\begin{subequations}\label{eq:time_dep_pb}
\begin{align}
- L^{-2}\partial_y^2\sigma+\sigma &= \mathcal{P} c , \label{eq:time_dep_pb_a} \\
\partial_t\left(L c\right)+\partial_y\left(w c\right) &= L^{-1}\partial_y^2 c , \label{eq:time_dep_pb_b} 
\end{align}
\end{subequations}
where Eq.~\eqref{eq:time_dep_pb_a} is just the dimensionless constitutive relation of the active gel combined 
with Eq.~\eqref{e:eq_stress2_b}, while Eq.~\eqref{eq:time_dep_pb_b} is the result of the application of the chain rule 
to the dimensionless form of Eq.~\eqref{e:motors_conservation}.
Here $w = v-\dot{G}-(y- 1/2)\dot{L}$ is the relative velocity, 
$G(t)=[l_-(t)+l_+(t)]/2$ is the position of the geometric center of the cell, 
$v=L^{-1}\partial_y\sigma$ is the velocity of the F-actin in the laboratory frame of reference,
and the dimensionless parameter $\P$ is defined in Eq.~\eqref{e:Pcal}. 
Note that now the stress $\sigma= \P c+L^{-1}\partial_yv$ does not contain the term $p_h$ which has been adsorbed into the homeostatic length~$L_0$ defined in Eq.~\eqref{e:L0}. The boundary conditions at $y=\left\lbrace 0,1\right\rbrace $ read
\begin{subequations}\label{eq:time_dep_bc}
\begin{align}
\sigma &= -\mathcal{K}( L-\L), \label{eq:time_dep_bc_a} \\
w &= 0, \label{eq:time_dep_bc_b} \\
\partial_y c &= 0 , \label{eq:time_dep_bc_c}
\end{align}
\end{subequations}
where the dimensionless parameters $\K$ and $\L$ are defined in Eqs.~\eqref{e:Kcal} and \eqref{e:Lcal}, respectively. 
The initial conditions can be chosen in the form $c(y,0)=c_i(y)$, $l_-(0)=l_-^i$, and $L(0)=L_i$.  

From Eqs.~\eqref{eq:time_dep_pb_a}, \eqref{eq:time_dep_bc_a}, and \eqref{eq:time_dep_bc_b} 
we can obtain the expressions for the cell speed \cite{Car_njp11,RecPutTru_prl13},
\begin{equation} 
\label{e:fvc1} \dot{G} =\frac{\mathcal{P} L}{2}\int_0^1\frac{\sinh[L(1/2-y)]}{\sinh(L/2)} \,c(y,t)\, dy,
\end{equation}
and for the rate of change of the  cell length, 
\begin{align}
\label{e:fvc2} \dot{L} &= -2\mathcal{K}\left( L-\L \right)\tanh\left(  L/2\right)  \nonumber \\ 
& \hspace{10ex} -\mathcal{P} L \int_0^1\frac{\cosh[L(1/2-y)]}{\cosh(L/2)} \,c(y,t)\, dy .
\end{align}
From Eq.~\eqref{e:fvc1}, we see that motility is associated with the emergence of an uneven motor distribution  and that  symmetrization of the motor distribution can lead to the cell arrest. In addition Eq.~\eqref{e:fvc2} shows that the steady-state length results from an interplay between the quasielastic, homeostatic resistance and the active shortening due to contractility. 

To summarize, the mechanism ensuring cell polarization in this model is based on the positive feedback exhibited by the Keller-Segel system \eqref{eq:time_dep_pb}: motor inhomogeneity generates gradients of contractile stress which in turn generate mass transport amplifying motor inhomogeneity. 
As a result motors localize on one side of the cell. When the corresponding cell boundary is not anchored, 
the whole system starts to move given that the symmetry of traction forces is broken. 
Diffusion can prevent such motors localization and therefore inhibit motility. However, as contraction builds up, 
the symmetric nonmotile state eventually loses stability. The externally applied mechanical stress, 
setting the \emph{homeostatic length} $\L$, controls the length of the cell, and hence, 
the ability of diffusion to suppress polarization. The next sections are devoted to quantifying the implied 
stress induced stabilization.

\section{Steady-state regimes}\label{s:steady}

Steady-state solutions of the system of Eqs.~\eqref{eq:time_dep_pb} and \eqref{eq:time_dep_bc} 
are traveling waves with both fronts moving at the same constant velocity, i.e., $V=\dot{G}$. 
In such states, the length of the cell is  fixed as $\dot{L}=0$ and $\partial_tc=0$. 
The system  \eqref{eq:time_dep_pb} reduces to 
\begin{subequations}\label{e:tw_ode}
\begin{align} 
v' &= (\sigma-\P  c)L , \label{e:tw_ode_a} \\
\sigma ' &= v L , \label{e:tw_ode_b} \\
c' &= (v-V) c L , \label{e:tw_ode_c}
\end{align}
\end{subequations}
where prime denotes the derivative with respect to $y$.  
The boundary conditions \eqref{eq:time_dep_bc} at $y=\{0,1\}$ now read
\be
\label{e:tw_ode_bc}
\sigma =-\mathcal{K}(L-\mathcal{L}),\, v=V .
\ee
Since the velocity $V$ and the length $L$ are to be determined self-consistently and that we are left with five unknown constants,
the four algebraic conditions \eqref{e:tw_ode_bc} must be supplemented by the constraint on 
the total mass of motors $L\int_0^1 c(y)dy=1$.

\emph{Homogeneous states.} The homogeneous solutions of Eqs.~\eqref{e:tw_ode} satisfy $v= 0$, $c = 1/L$, and $\sigma= -\mathcal{K}(L-\L)$. This class of solutions  corresponds to stationary states with $V=0$. The length of the cell, $L$, is determined by the quadratic equation \mbox{$L^2 -\L L +\mathcal{P}/\mathcal{K}=0$}, which follows from the condition $\sigma=\P c$. Provided that $\P\leq\K \L^2/4,$  the cell has two trivial configurations, 
\be
\hat{L}_{\pm}\big(\L,\mathcal{P} /\mathcal{K}\big) =   \big(\L \pm \sqrt{\L^2-4\mathcal{P} /\mathcal{K}}\big)/2, 
\label{e:Lpm}
\ee
merging at 
\be
\label{collapse_cond}
\L_c = 2\sqrt{\mathcal{P} /\mathcal{K}}, L_c = \sqrt{\P/\K}.
\ee

\emph{Collapsed states.} At $\P>\K \L^2/4$ the homogeneous solutions do not exist because the quasielastic resistance 
is not buttressed sufficiently by the external stretching to resist the contraction. The possibility of contraction-induced cell collapse can also be seen in the vertex model setting~\cite{lin2017activation}. 

To understand more clearly the ensuing singular behavior we need to look at the transient dynamics leading to the cell collapse. The asymptotic behavior of the solution of Eqs.~\eqref{eq:time_dep_pb} and~\eqref{eq:time_dep_bc} when the length of the cell, $L$, is vanishingly small can be represented in the form 
\begin{align*}
c(y,t) &= c_{-1}(y)L^{-1}(t)+c_0(y) L^{0}(t)+\ldots\\
v(y,t) &= v_{-1}(y)L^{-1}(t)+v_0(y) L^{0}(t)+\ldots\\
\sigma(y,t) &= \sigma_{-1}(y)L^{-1}(t)+\sigma_0(y) L^{0}(t)+\ldots
\end{align*}
The substitution of these expansions into the equations gives 
\begin{align*}
c_{-1}(y) &= 1 ,  \\
v_{-1}(y) &= 0 , \; v_0(y)=\mathcal{P}  ( 1/2-y) , \\
\sigma_{-1}(y) &= 0,  \; \sigma_0(y)=\mathcal{K} \L .
\end{align*}
In addition we obtain $\dot{L}(t)=-\mathcal{P}$ and $\dot{G}(t)=0$, which means that a static cell segment collapses 
in finite time $O(\P^{-1})$.   
In the process, the motor concentration diverges while remaining spatially homogeneous. 

The singular behavior of the solution signifies the failure of some of the model assumptions and it is necessary to find a physically informed regularization of the model which removes the singularity. The two natural paths are to reinforce the  length-regulating mechanism and/or to saturate the activity of motors at large concentrations.

Focusing first on the length-regulating mechanism we may assume that the effective spring accounting for the global homeostatic feedback, is nonlinear. Then Eq.~\eqref{e:bc_stress} can be rewritten in the form
\be\label{e:bc_stress1}
\eta \partial_xv(l_\pm(t),t)+\chi c(l_\pm(t),t)=-k\big[L-L_0f(L)\big],
\ee
where $f(L)$ is a function approaching $1$ as $L\to\infty$ and diverging as $L \to 0$. A simple choice ensuring the desired behavior is $f(L)=1+(L_f/L)^\alpha$, where $L_f$ is a new characteristic length and $\alpha$ is a phenomenological exponent. A physical motivation for the choice of $\alpha=2$ is presented in Appendix~\ref{sec:appendix_A}. 

We can also account for a size-dependent motor contractility which mimics the effects of crowding-related frustration at small cell lengths. To this end we can replace $\chi$ by $\chi g(L)$, where, for instance, $1/g(L)= 1+(L_g/L)^\beta$ with $L_g$ being another characteristic length and $\beta$ another phenomenological exponent. We may also assume that the total number of attached motors is biochemically regulated at the global level \cite{Noll2017} so that it decreases with the cell length. In this case, we should replace $M$ by $M h(L)$, where, for instance, $1/h(L)= 1+(L_h/L)^\gamma$ with $L_h$, yet another characteristic length, and $\gamma$, another phenomenological exponent. 
\begin{figure}
\centering
\includegraphics[scale=0.9]{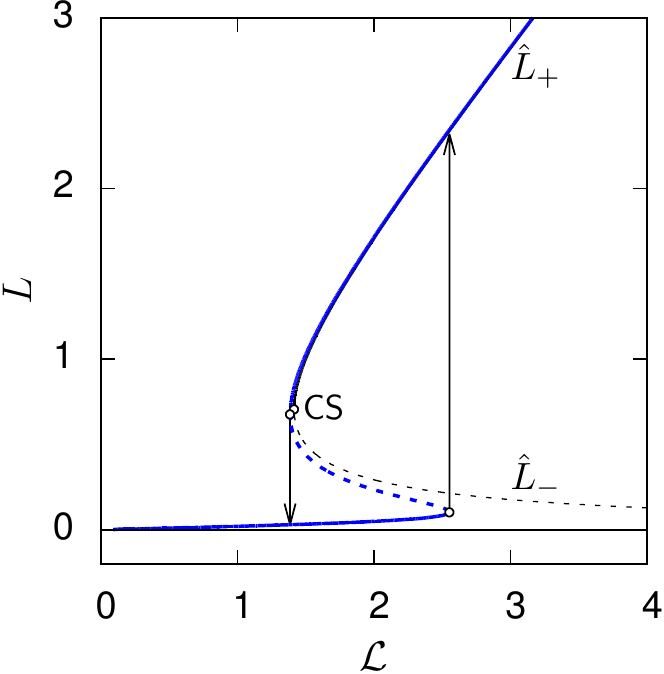}
\caption{Length of the homogeneous cell as a function of~$\L$ in the nonregularized case (thin black line) and in the regularized case (thick blue line). The horizontal solid black line shows collapsed solutions in the nonregularized model. The chosen regularization corresponds to $g=h=1$, $L_f=0.1$, and $\alpha=2$. Parameter: $\P/\K=0.5$.}  
\label{f:lengthvssqueeze}
\end{figure}

Note that all these sigmoidal dependencies belong to the class of Hill-Langmuir  equations with the coefficients $\alpha$, $\beta$, and $\gamma$ usually quantifying the degree of cooperativity. They must be non-negative and sufficiently large to prevent the collapse of a cell. More precisely, since the cell length can be found from the dimensionless algebraic equation $\P h(L)g(L)=-L\mathcal{K}[L-\L f(L)]$, we can conclude that for $L$ much larger than any of the regularizing lengths, the solution coincides with Eq.~\eqref{e:Lpm}, but for sufficiently small $L$ and $\alpha+\beta+\gamma>1$ there will be another branch $\hat{L} =(\mathcal{K}\L/\P) (L_f^{\alpha}L_g^{\beta}L_h^{\gamma})^{1/\nu}$, where $\nu=\alpha+\beta+\gamma-1$. 
The associated velocity, stress, and motor concentration can then be found from the relations $\hat{\sigma}=\mathcal{K} L \left( L_f/L_r \right)^\alpha$, $\hat{v}=0$ and $\hat{c}= L_r^{\gamma-1}/L_h^\gamma$.  

An example of the coexistence of the two branches is presented in Fig.~\ref{f:lengthvssqueeze} (thick blue line). As we see, the collapse of the cell induced by squeezing takes place discontinuously near the critical value of $\L$ given by Eq.~\eqref{collapse_cond}. Interestingly, the regularized theory also predicts an inverse transition from collapsed to noncollapsed state when $\L$ increases. The ensuing hysteresis is reminiscent of what happens in the three-dimensional theory of cytokinesis~\cite{turlier2014furrow}, which suggests that a possible interpretation of the collapsed states could be associated with cell division. The ambiguity, however, remains since our prototypical model cannot really distinguish between various modalities of the abrupt loss of cell integrity; in particular, it can confuse cell death with cell division. 
Moreover, the  one-dimensional model cannot really differentiate a change of cell volume from a change of cell
shape. So, it is not clear that collapsed solutions should be interpreted as cells which fully lose their volume. They rather
indicate that the cell cortex undergoes a drastic singularity involved, for instance, in a transition from a squamous to 
columnar state~\cite{hannezo2014theory}.

\emph{Inhomogeneous steady states.} Nonsingular solutions of Eqs.~\eqref{e:tw_ode} can be of \emph{motile} or \emph{static} type depending on the symmetry of the motor distribution [see Eq.~\eqref{e:fvc1}]. The stability analysis presented in Sec.~\ref{s:numstability} suggests that the only stable inhomogeneous solutions are nonsymmetric motile states with myosin motors localized at the trailing edge; in view of the reflectional symmetry of the problem such solutions appear in pairs. Below we show that an asymptotic representation of such motile solutions can be computed analytically in the double limit $\mathcal{P}\rightarrow\infty$ and $\mathcal{K}\rightarrow\infty $ with the ratio $r=\mathcal{P}/\mathcal{K}$ remaining finite. 
 
Suppose that one of the twin configurations $c(y)$, solving Eqs.~\eqref{e:tw_ode}, has a maximum at $y=0$ and decays to zero away from this point. Then the asymptotic representation of the solution can be written in the form (see Appendix~\ref{sec:appendix_B}) 
\be
c(y)=\frac{\P}{2\cosh^2(\P L y/2)},
\label{e:c_asymp}
\ee
where $L$ solves the algebraic equation
\be 
\label{e:collapsed1}
(\P/2) \coth(L/2) =-\mathcal{K} (L-\L).
\ee
The obtained result shows that the motors distribution becomes infinitely localized in the limit \mbox{$\P\rightarrow\infty$}. If we also assume that $\P\sim r\K $ where $r$ is finite we obtain that Eq.~\eqref{e:collapsed1} has solutions if and only if 
\begin{equation}\label{e:collapsed_anal_large_lambda}
(r/2) \sqrt{4/r+1} \leq \L - 2\sinh^{-1}(\sqrt{r}/2).
\end{equation} 
Inequality \eqref{e:collapsed_anal_large_lambda} represents the collapse condition and, from Eq.~\eqref{e:collapsed1}, we see that the collapse of a motile cell takes place at finite length $L=2 \sinh ^{-1}(\sqrt{r}/2)$. In the limit $\L r\ll 1$, condition~\eqref{e:collapsed_anal_large_lambda} takes a particularly simple form 
\be
\label{e:simplified_collapsed_anal_large_lambda}
\P \leq \K\L^2/4 .
\ee
Observe that Eq.~\eqref{e:simplified_collapsed_anal_large_lambda} coincides with Eq.~\eqref{collapse_cond}, which suggests that the inhomogeneity of motor concentration affects only weakly the onset of the transition from regular to singular regimes. 
\begin{figure}
\centering
\includegraphics[scale=0.65]{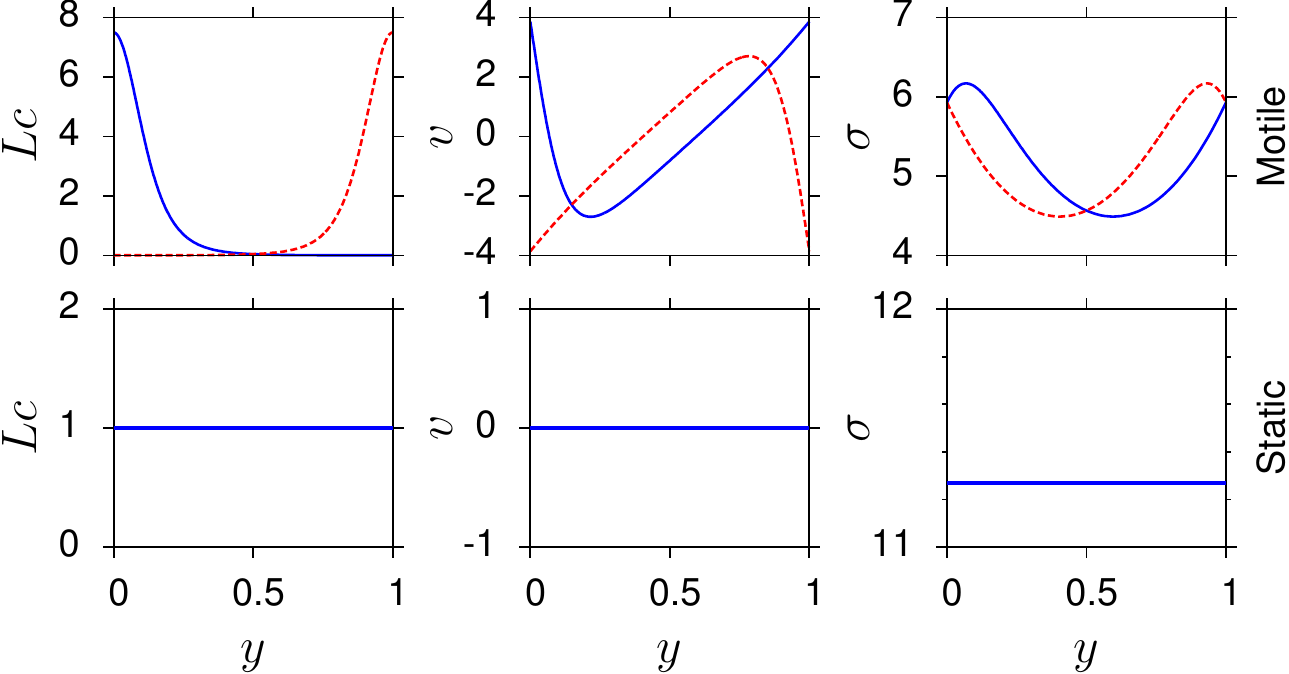}
\caption{Solution profiles for $\P=10$ and two values of $\L$. The motile solutions correspond to $\L=2$ and the homogeneous stable static state to $\L=1$. For the motile solutions, whether the cell velocity is positive (motors are localized near the boundary $y=0$, solid blue curve) or negative (motors are localized near the boundary $y=1$, dashed red curve) depends on the initial noise~$\zeta$; however, both configurations are equiprobable; $V\approx \pm 3.86$, $L\approx 1.94$.
Parameter: $\K=100$.}
\label{f:profiles}
\end{figure}
 
\begin{figure}
\centering
\includegraphics[scale=0.8]{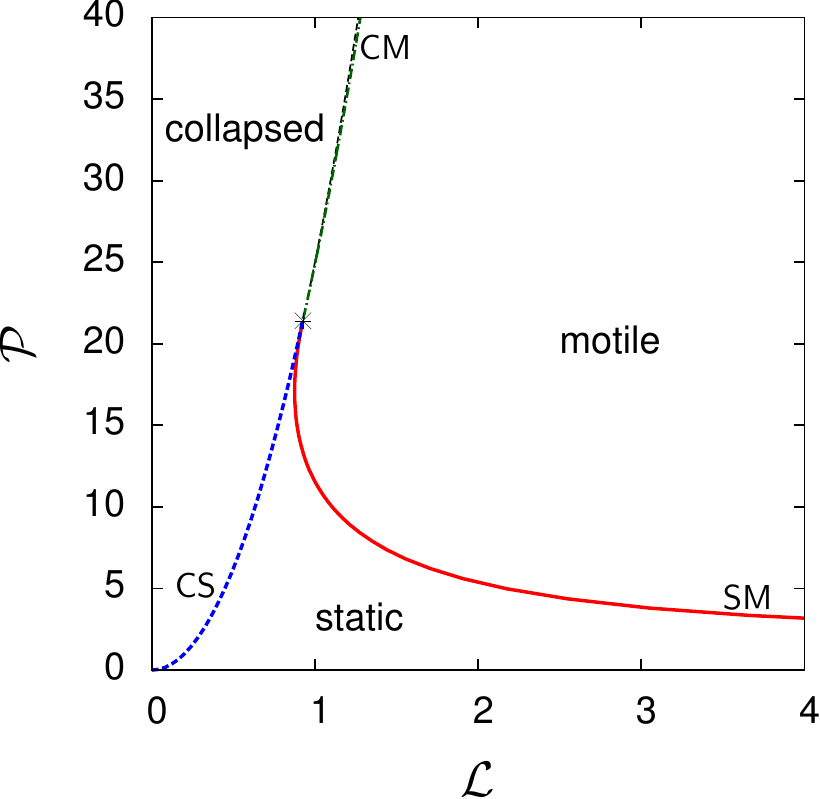}
\caption{Typical phase diagram in the parameter plane~$(\L,\P)$, where $\L$ is the homeostatic length, which can be interpreted as a scaled stress, and $\P$ is the dimensionless measure of the myosin contractility. The three reachable configurations are \emph{motile}~({\sf M}), \emph{static}~({\sf S}), and \emph{collapsed} ({\sf C}), given the initial data indicated in the text. The dashed blue and solid red lines correspond to {\sf CS} and {\sf SM} thresholds, respectively. The black dashed line, essentially overlapping with the green dash-dotted line, indicates the approximation of the {\sf CM} threshold given by Eq.~\eqref{e:simplified_collapsed_anal_large_lambda}. Parameter: $\K=100$.}
\label{f:phase_diagram}
\end{figure}

\begin{figure*}
\centering
\includegraphics[scale=0.36]{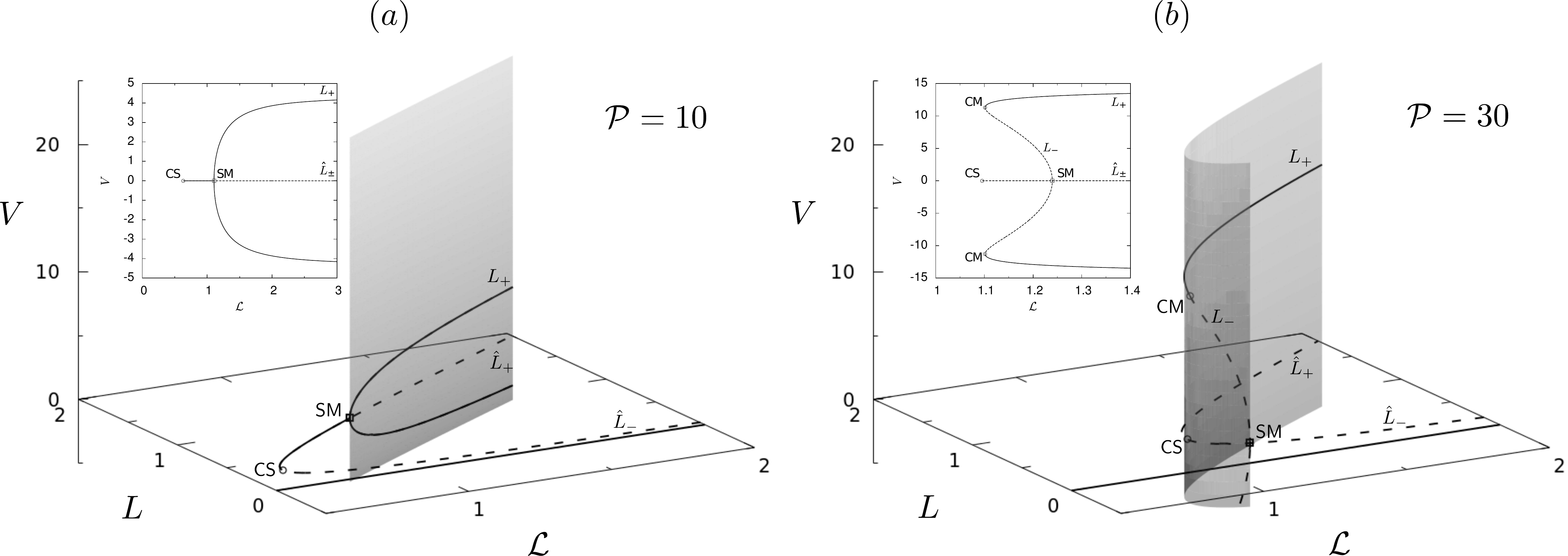}
\caption{
Typical bifurcation diagrams with the homeostatic length $\L$ chosen as a parameter: (a) below the critical end point $\mathcal{P}=10<\mathcal{P}_*$ and (b) above the critical end point $\mathcal{P}=30>\mathcal{P}_*$. Solution branches representing stable attractors in the initial value problem [Eqs.~\eqref{eq:time_dep_pb} and \eqref{eq:time_dep_bc}] are shown by  solid lines; locally unstable solutions are presented by dotted lines. Insets represent  projections of the three-dimensional diagram on the $(\L,V)$ parameter plane. 
Parameter: $\K=100$.}
\label{f:bfn_diag}
\end{figure*}

\section{Phase diagram}\label{s:numstability}

To identify stable traveling wave solutions we solved numerically a set of initial value problems for the original nonsteady system of Eqs.~\eqref{eq:time_dep_pb} and \eqref{eq:time_dep_bc}. We fixed parameter $\mathcal{K}$, whose effect on the behavior of the system has been studied earlier \cite{RecPutTru_jmps15}, and focused on the parameter space $(\L,\P)$. As initial data we used $L_i=\L$ and $c_i(y)=1/\L+ \zeta(y)$ where $\zeta$ is a  random perturbation with zero average. We integrated Eqs.~\eqref{eq:time_dep_pb} and \eqref{eq:time_dep_bc} numerically, analysing transients as the system approached one of the steady states (see \cite{RecPutTru_jmps15} for the method). 
 
The outcomes of our numerical experiments fell into three categories: 
(i)~convergence to a motile solution with motors localized at the trailing edge ({\sf M}, top row in Fig.~\ref{f:profiles}); 
(ii)~convergence to a homogeneous static solution of the $\hat{L}_+$  type ({\sf S}, bottom row in Fig.~\ref{f:profiles}); 
and  (iii)~collapsing solution approaching a singular state {\sf C} in a finite time. 
The stability domains of these three ``phases'' within the $(\L,\P)$-plane are shown in Fig.~\ref{f:phase_diagram}.

\emph{Stability thresholds.} The motile ({\sf M}), static ({\sf S}), and collapsed ({\sf C}) regimes are separated by three boundaries which meet at a triple point (symbol~$*$ in Fig.~\ref{f:phase_diagram}): the line {\sf CS} separates collapsed and static ($\hat{L}_+$) solutions, the line {\sf SM} separates static and motile solutions, and the line {\sf CM} separates collapsed and motile solutions.

The {\sf CS} line is captured by condition \eqref{collapse_cond} because beyond this line static homogeneous solutions cease to exist. The change of the regime here is abrupt and can be interpreted as a \emph{first order transition}.

The {\sf SM} line can be also described analytically if we linearize Eqs.~\eqref{eq:time_dep_pb} and \eqref{eq:time_dep_bc} around the homogeneous solution $\hat{L}_+$ and study the limits of linear stability of the homogeneous solution. A standard analysis \citep{RecPutTru_prl13,RecPutTru_jmps15} gives the instability condition $\tanh\left( \omega/2\right) \hat{L}_+= \mathcal{P}\omega/2$, where $\omega=(\hat{L}_+^2-\mathcal{P} \hat{L}_+ )^{1/2}$. The bifurcation is of pitchfork type and can be interpreted as a \emph{second order phase transition}. Note that a mild generalization of the model accounting for the saturation of contractility as a function of the motor concentration turns the supercritical bifurcation into a subcritical bifurcation~\cite{RecPutTru_jmps15}.

The {\sf CM} line, separating motile and collapsed states and corresponding to another \emph{first order transition}, cannot be expressed analytically. However, at large values of $\P$ and $\K$ such that $\K\L/\P\ll 1$, this transition is asymptotically captured by condition \eqref{e:collapsed_anal_large_lambda}. For the realistic parameters used in Fig.~\ref{f:phase_diagram}, expression \eqref{e:collapsed_anal_large_lambda} provides a remarkably good approximation for the {\sf CM} line starting already at the triple point.  

The coordinates of the triple point $(\L_*,\P_*)$, which in view of the nature of the {\sf SM}, {\sf CM}, and {\sf CS} transitions should be rather called the ``critical end point,'' can be found as the intersection of the {\sf CS} and {\sf SM} lines. Therefore, $\mathcal{P} _*=\mathcal{K} \mathcal{L}_*^2/4,$ where $\mathcal{L}^*$ satisfies the equation $2\tanh(\omega_*/2)=\mathcal{K} \omega_*\mathcal{L}_*^2/4$ with $\omega_*^2=\mathcal{L}_*^2(\mathcal{K}  \mathcal{L}_*^2-2)/8$.  

\emph{Bifurcation patterns.} To understand better the relation between different types of traveling wave solutions in this problem, we used a numerical continuation method~\cite{DoeKelKer_ijbc91} with parameter $\L$ varying around the critical end point. We found that, in full agreement with the analysis above, the bifurcation diagrams can be of two types depending on the value of the parameter $\P$ (see Fig.~\ref{f:bfn_diag}).

At $\mathcal{P}<\mathcal{P}_*$ the motile solution bifurcates from the homogeneous solution $\hat{L}_+$ through a supercritical pitchfork, which we interpret as a continuous phase transition at a critical value of the parameter $\L$ given by the instability condition. 
We recall that the locus of these points in the parameter space $(\L,\P)$ is represented in Fig.~\ref{f:phase_diagram} by the solid red line, {\sf SM}. 

\begin{figure}
\centering
\includegraphics[scale=0.35]{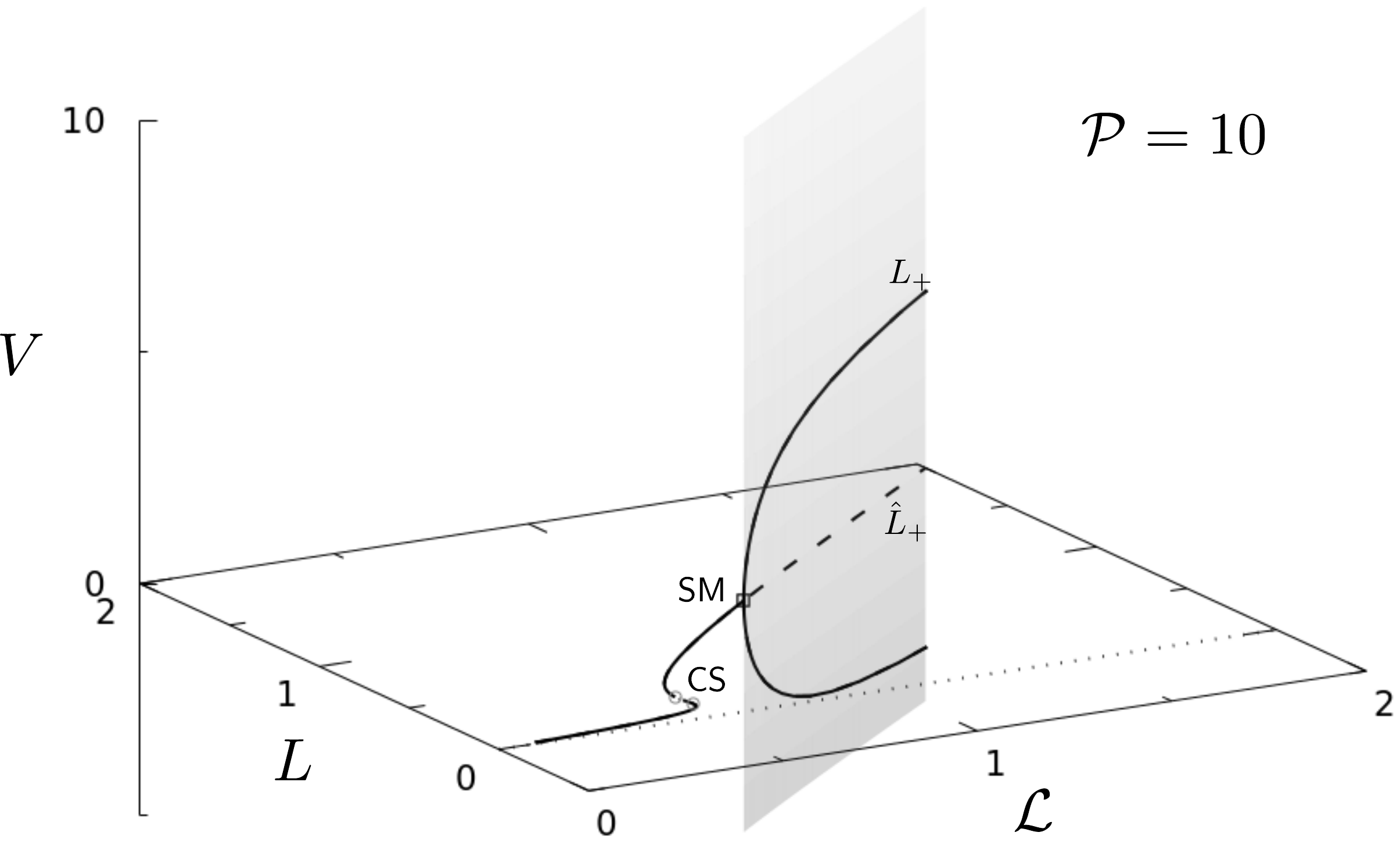}
\caption{An example of a bifurcation diagram for the regularized model with $g=h=1$, $L_f=0.1$, and $\alpha=2$.
Parameters: $\P=10$ and $\K=100$. }
\label{f:bd_reg}
\end{figure}

While the degenerate collapsed solutions are not reachable by the continuation method, we also show in Fig.~\ref{f:bfn_diag} the collapsed solutions with zero length. Observe that those are the only possible solutions if the parameter $\L$ is below the turning point {\sf CS}. The locus of the turning points {\sf CS} is shown in Fig.~\ref{f:phase_diagram} by a dashed blue line given by Eq.~\eqref{collapse_cond}.   

The collapsed solutions are more adequately represented by the regularized models discussed in Sec.~\ref{s:steady}. In Fig.~ \ref{f:bd_reg} we show a regularization-induced modification of the bifurcation diagram presented in Fig.~\ref{f:bfn_diag}(a) for the system  \eqref{eq:time_dep_pb} with $g(L)=h(L)=1$ and  $f(L) \ne 1$. A comparison of Figs.~\ref{f:bd_reg} and~\ref{f:bfn_diag}(a) shows that regularization did not affect significantly the bifurcation from static to motile solutions even though now there is a finite coexistence domain between collapsed and noncollapsed static solutions (illustrated in more detail in Fig. ~\ref{f:lengthvssqueeze}).

Returning to the original, nonregularized setting, we now consider the second type of behavior of the system corresponding to the range $\P>\P_*$, where the motile solutions bifurcate from the unstable branch of the homogeneous static solutions $\hat{L}_-$. Along the bifurcated motile branch, there is now a turning point {\sf CM} where the stability of the motile regimes in the full time-dependent problem, Eqs.~\eqref{eq:time_dep_pb} and \eqref{eq:time_dep_bc}, is lost. If~$\L$ is decreased beyond this turning point, there are no more nondegenerate stable static solutions and, this is the reason why our numerical experiments showed in this range a discontinuous transition to the collapsed state [see Fig.~\ref{f:bfn_diag}(b)]. The locus of the {\sf CM} turning points in the parameter space $(\L,\P)$ is shown in Fig.~\ref{f:phase_diagram} by a dashed green line. 
  
\emph{Sensitivity to initial data.} Since our numerical study of the time-dependent problem [Eqs.~\eqref{eq:time_dep_pb} and \eqref{eq:time_dep_bc}] was necessarily limited in scope, the actual basin of attraction of each of our three main regimes (static, motile, and collapsed) was not fully mapped in the space of parameters describing the  initial data. We have found, however, that the phase diagram in Fig.~\ref{f:phase_diagram} is robust with respect to rather general perturbations of the initial concentration field.  

The dependence of the results on the choice of the initial length is more sensitive. For instance, choosing the initial length in the range $L_i\leq \hat{L}_-$ leads to a collapse independently of the value of $\mathcal{P}$ and $\mathcal{L}$. This empirical result can be supported by the observation, that for configurations with homogeneous distribution of myosin, Eq.~\ref{e:fvc2}, reduces to  
\be
\dot{L} = -2\mathcal{K} \tanh(L/2)(L-\hat{L}_-)(L-\hat{L}_+)/L .
\ee
This equation reaffirms that the homogeneous state $\hat{L}_+$ is  stable with the basin $]\hat{L}_{-},\infty [$, while the state $\hat{L}_-$ is unstable and  any initial state with $[0,\hat{L}_-[$ should  unconditionally collapse.

\section{Discussion}\label{s:discussion}

Observe that the active mechanism in our model, whose strength scales with parameter $\P$ defined by Eq.~\eqref{e:Pcal}, is responsible for two different physical effects. 

On the one hand, an increase of $\P$ can lead to a collapse of a homogeneous configuration of a cell. This implies breaking the balance between motor contractility and the effective quasielastic resistance. The transition is discontinuous (first order) even though appropriate regularization of the problem can prevent the formation of a singularity.

On the other hand, an increase of $\P$ can be also responsible for the loss of homogeneity and the emergence of polarity. If the incipient concentration gradients are not  suppressed by diffusion, motors localize and the  cell starts to move. The threshold of such instability is sensitive to the homeostatic length~$\L$, defined by Eq.~\eqref{e:Lcal}, because the homogenizing effect of diffusion is more potent in a small domain than in a large domain. In contrast to the collapse transition, the motility transition is continuous (second order).

In view of the above, the emergence of the three main configurations---\emph{motile}, \emph{static}, and \emph{collapsed}---becomes natural. At $\P<\P_*$ the homogeneous solution $\hat{L}_+$ is replaced by a motile solution at sufficiently large value of $\L$ when the diffusion is no longer sufficient to prevent the contraction-induced drift. On the contrary, when $\L$ decreases below sufficiently small value, the homogeneous contraction overcomes the cell elastic resistance which triggers the collapse. In between, there is a finite interval of sufficiently small values of $\L$, where stable static solutions exist. In the $\P>\P_*$ range, both the symmetrization and the collapse take place simultaneously at a single threshold. The main conclusion is that stable nontrivial static configurations are delicate and may be achievable only in the presence of stretching force dipoles applied to the cell from the environment.  

We now interpret these findings in the context of tissue homeostasis. Cells in tissues are typically subjected to stresses exerted by their neighbors and mediated by adherens junctions~\cite{guillot2013mechanics}. For instance, when epithelial cells reach confluence, they become caged by their neighbors and the formation of adherens junctions effectively jams the cell monolayer through a process known as contact inhibition~\cite{garcia2015physics}. The tension in such monolayers has been monitored using atomic force microscopy indentation~\cite{harris2014formation} and, during the initial formation of the monolayer when E-cadherin clusters are formed between the cells that are still polarized, the tension and the surface area of the cells increase indicating a rise of the homeostatic length. After this initial stage, the monolayer tension significantly drops. If we interpret this drop as a decrease of $\L$, the observed  cell symmetrization within a monolayer becomes natural. Interestingly, the predicted transition from symmetric to collapsed state at even smaller values of $\L$ has been also observed in monolayers. As explained in \cite{wyatt2016question}, local decrease of the tensile stress in a tissue can lead to the increase of tissue density (i.e., smaller homeostatic length), which triggers cell extrusion events. The latter may be associated with our cell collapse (see also~\cite{marinari2012live, eisenhoffer2012crowding, gudipaty2017mechanical}). 
 
It is also known that scratch-wound assays and laser ablations \cite{Tam_nmat11,weber2012mechanoresponsive,vedula2013collective} reduce the compressive stresses  by creating an available free space. Our model shows that such a reduction could be sufficient for the spontaneous initiation of motility of the leader cells at the margin of the tissue. This puts the tissue under tension~\cite{vincent2015active,blanch2017effective} and further increases bulk motility needed to heal the wound. Chemotactic signaling is clearly also necessary for normal wound closure.  However, the importance of mechanics is corroborated by the observation that a moderate external cell stretching accelerates wound closure~\cite{toume2017low}.
   
The closely related phenomenon of the EMT is of significant interest because of its central role in diseases such as fibrosis and cancer \cite{thiery2009epithelial}. It is known that the mechanosensitive activation pathways of EMT involve both exogenous and endogenous stresses supported by the cytoskeleton (see~\cite{gjorevski2012regulation} and references therein). If an external stress is applied to a cell in a tissue, the parameter $\L$ will increase,  leading to the mechanical loss of stability of a static configuration. The ensuing F-actin flow is characterized by a polymerization of the F-actin meshwork at the cell front. Such a polymerization reduces the pool of G-actin with which a transcription factor MRTF-A is normally  associated. This transcription factor then becomes free to accumulate in the cell nucleus, where it promotes the expression of the genes regulating the disassembly of cell-cell contacts, which ultimately liberates the cell from the entanglement with its neighbors. 

While a direct pharmaceutical activation of the Rho pathway, which is known to affect the remodeling of the cytoskeleton, can also trigger the nuclear translocation of MRTF-A, our model suggests that both, internal and external stress creation, can generate alternative mechanical pathways regulating the polarization of cells and controlling in this way the emergence of EMT.

\section{Conclusions}\label{s:conclusion}

Cell motility is affected by externally applied forces. Here we argue that not only the resultant, but also the distribution of applied forces matters, in particular, that cell migration may be sensitive to mechanically balanced force couples with zero resultant. This may sound surprising since such couples can squeeze or stretch the cell but do not impose explicit directionality. 

To study the effect of the applied force couples, we employed a one-dimensional model incorporating active myosin contraction and accounting for the induced actin turnover. We showed that a symmetrization and a spontaneous motility arrest of a polarized cell can result from squeezing of a steadily moving cell. Conversely, we showed that, if squeezing is relaxed, a static symmetric cell may get again polarized and start to move. An analytical study of the model revealed the exact amount of mechanical stress needed to polarize a static cell and arrest a moving cell. 

The proposed model predicts further that sufficiently strong squeezing can lead to the loss of cell integrity and discontinuous collapse of the cell to (almost) zero length. We argued  that possible interpretations of the collapsed states could be associated with cell division or death.

The obtained phase diagram in the space of nondimensional parameters distinguishes between motile, static, and collapsed configurations of the cell. It can be relevant for the study of the~EMT and may help to understand  the mechanical aspects  of~CI. For instance, the model suggests that an increased cell tension may lead to cell polarization emulating~EMT. Instead, squeezing can be the origin of polarity loss, which is a characteristic feature of the CI phenomenon.  

Interestingly, our model reveals that self-induced squeezing associated with the active cortical contraction can lead to symmetrization and cell arrest even in the absence of the confining environment. Another insight is that internal homeostatic pressure, originating from actin turnover, may act quantitatively similar to the externally applied stress as a factor promoting cell polarization and motility. 

A shortcoming of our analytically tractable model is that neither static solutions nor regularized collapsed solutions carry a nontrivial F-actin flow. Such flows, however, can be recovered in this framework if we assume that contractility is space dependent and controlled by independent biochemical pathways \cite{turlier2014furrow}. Another weakness of the model is the neglect of protrusive stresses at the leading edge associated with actin polymerization. We also left aside the important question of the interplay between the resultant force and the applied force couple, which will manifest itself in a stress dependence of the force-velocity relations. All these issues require further research.

Despite these limitations, we expect that the transparency of the proposed model will motivate focused experimental investigations of the predicted  transitions between different regimes and inspire the development of biomimetic devices imitating the rich mechanical behavior of crawling cells. Indeed, by accounting for the whole spectrum of external loadings, the model bridges cellular and tissue scales, and opens new perspectives in the design of the  artificial analogs of collective cellular motility.

\section*{Acknowledgements}
T.P. acknowledges support from the EPSRC Engineering Nonlinearity project EP/K003836/1 and Norbert Hoffmann. 
L.T. was supported by the French governement under Grant No. ANR-10-IDEX-0001-02 PSL.

\appendix
 
\section{}\label{sec:appendix_A}

We represent a cell by a slab of height~$\h$ and length~$L$, confined to a track of constant width~$\w$. 
The slab is modeled as a solution of biopolymers enclosed in a plasmic membrane which adheres to the track. 

The free energy of the system, $\FE$, is the sum of the surface energy $\FE_s$ and a confinement energy $\FE_c$ (see~\cite{hannezo2014theory}). The surface energy can be written in the form $\FE_s=2\gamma_m(L\w+L\h+\w\h)-\gamma_sL\w$, where $-\gamma_s<0$  is the cell-substrate surface tension showing that spreading is energetically favored and $\gamma_m>0$ is the conventional surface tension of the free membrane. The confinement energy is assumed to be of entropic nature, accounting for the presence of Gaussian polymers inside the slab. The simplest expression then is $\FE_c=\A(1/\h^2+1/\w^2+1/L^2)$, where~$\A$ is a rheological coefficient whose value depends on the nature of the polymers~\cite{DeGennes1979}.

Suppose next that, while the surface area of the cell can change through the addition of lipids to the membrane~\cite{keren2011cell}, the cell volume~$\V$ remains constant, tightly regulated by an active osmotic balance between the cell interior and exterior~\cite{jiang2013cellular, hui2014volumetric, mcgrail2015osmotic}. Using the geometrical constraint $\V=L\h\w$, we then eliminate height~$\h$ from the expression of the free energy to obtain 
$\FE=-\gamma_s L\w + 2\gamma_m (L\w+\V/\w+\V/L)  + \A [(L\w/\V)^2+1/\w^2+1/L^2 ].$ 

The internal traction $\tau =-(\h\w)^{-1}\partial_L \FE$ takes the form 
\be\small
\tau = - \frac{\gamma L}{\tilde{L}_f^2\tilde{L}_0}\left\lbrace L \left[1-\left(\frac{\tilde{L}_f}{L}\right)^4\right] - \tilde{L}_0\left[1+ \frac{\gamma_m}{\gamma}\left(\frac{\tilde{L}_f}{L}\right)^2\right] \right\rbrace ,
\label{e:tau_int_exact}
\ee
where $\tilde{L}_f=\sqrt{\V/\w}$, $\tilde{L}_0=\gamma\V^2/(2\A\w)$, and $\gamma=\gamma_s-2\gamma_m>0$. In the regime $L\gg\h$, describing cell spreading, 
we can drop the fourth power term in Eq.~\eqref{e:tau_int_exact}. If we further assume that $L\sim\tilde{L}_0$, we obtain Eq.~\eqref{e:bc_stress1} with 
$\alpha=2$; however, our main conclusions about the effect of regularization will survive even without this assumption. 

\section{}\label{sec:appendix_B}

To obtain the desired asymptotic representations of inhomogeneous solutions of Eqs.~\eqref{e:tw_ode}, it will be convenient to reformulate Eqs.~\eqref{e:tw_ode} as a nonlinear integral equation. First, by combining Eqs.~\eqref{e:tw_ode_b} and \eqref{e:tw_ode_c}, the motor concentration is expressed in the form
\begin{equation}
\label{e:concentration_stress}
c(y)=\frac{1}{L}\frac{e^{\Psi(y)}}{\int_0^1 e^{\Psi(y)}dy} ,
\end{equation}
where $\Psi(y)=\sigma(y)-VLy$.  In turn, the stress field is written as
\begin{equation}
\label{e:concentration_stress1}
\sigma(y)=\mathcal{P}  L \int_0^1\phi(y,z)c(z)dz+\bar{\sigma}\theta(y) ,
\end{equation}
where the stress on the boundary is $\bar{\sigma}=-\K(L-\L)$.  
Introducing the Heaviside function $H(x)$, we express the auxiliary functions in Eq.~\eqref{e:concentration_stress1} as 
\begin{align*}
\phi(y,z) =& \frac{\sinh[L(1-z)]\sinh(Ly)}{\sinh(L)}\\ 
          &-H(y-z)\sinh[L(y-z)] ,\\
\theta(y) =& \frac{\cosh(L y)[1 - \cosh(L) + \sinh(L)]}{\sinh(L)} .
\end{align*}
Using the boundary conditions, we can now link the cell velocity $V$ and its length $L$ to the concentration field with 
\begin{subequations}\label{e:velocity_stress_integ}
\begin{align}
V &= \mathcal{P} L\int_0^1\alpha(y)c(y)dy, \label{e:velocity_stress_integ_a} \\ 
\bar{\sigma} &= \mathcal{P}  L\int_0^1\beta(y)c(y)dy, \label{e:velocity_stress_integ_b}
\end{align}
\end{subequations}
where  
$$
\alpha(y)=\frac{\sinh[L( 1/2-y)]}{2\sinh(L/2)}, \;
\beta(y)=\frac{\cosh[L(1/2-y)]}{2\sinh(L/2)}.
$$
Finally, collecting all these expressions together, we obtain an integral representation for $\Psi(y)$:
\begin{equation}\label{e:integral}
\Psi(y)=\mathcal{P}  L \int_0^1\left[ \phi(y,z)+\theta(z)\beta(z)-Ly\alpha(z)\right] c(z)dz .
\end{equation}
Substituting Eq.~\eqref{e:integral} into Eq.~\eqref{e:concentration_stress} we obtain the desired integral equation for $c(y)$
$$
c(y)=\frac{1}{L}\frac{e^{\mathcal{P}  L \int_0^1\left[ \phi(y,z)+\theta(z)\beta(z)-Ly\alpha(z)\right] c(z)dz}}{\int_0^1 e^{\mathcal{P}  L \int_0^1\left[ \phi(s,z)+\theta(z)\beta(z)-Ls\alpha(z)\right] c(z)dz}ds}.
$$
Next, as the main contribution to the integral in Eq.~\eqref{e:integral} arises from the fast decay of the 
concentration field in the boundary layer at the rear front, we can use the method of matched asymptotic expansions, 
e.g.,~\cite{Hinch1991,Verhulst2005}, to obtain the asymptotic analog of Eq.~\eqref{e:integral} for $\P\gg 1$.
After rescaling the concentration field and writing it as $\tilde{c}(u)=\mathcal{P}^{-1}c(u/\mathcal{P})$, 
where we introduced the blow-up spatial coordinate $u\in [0,\mathcal{P}]$, we obtain
\be
\label{e:integral1a}
\Psi(u) \simeq L \int_0^{\infty} R(u,v)\tilde{c}(v)dv + \text{const},
\ee
with the kernel 
$$
R(u,v)=\left\lbrace 
\begin{array}{lc}
\frac{L}{2}(v-2u) & \text{if}\; u\leq v, \\
-\frac{L}{2}v & \text{if}\; u> v.
\end{array}\right.
$$
Note that the constant in Eq.~\eqref{e:integral1a} is irrelevant because it does not affect Eq.~\eqref{e:concentration_stress}.

We now reformulate the integral equation \eqref{e:concentration_stress} in a simpler form
\be
\label{e:integral1b}
\tilde{c}(u)=\frac{e^{L^2\Phi(u)}}{L \int_0^{\infty} e^{L^2\Phi(v)}dv} , \; 
\Phi(u)=\int_0^u(v-u)\tilde{c}(v)dv.
\ee
The remaining parameter $r=\P/\K$ enters these relations indirectly through the cell length $L$ which is still unknown.

Given that $\Phi'(u)=-\int_0^u\tilde{c}(v)dv $ and $\Phi''(u)=-\tilde{c}(u)$, we then rewrite Eq.~\eqref{e:integral1b} in the form
\begin{equation}
\label{e:integral3}
L\Phi''(u)\int_0^\infty e^{L^2\Phi(v)}dv+ e^{L^2\Phi(u)} = 0.
\end{equation}
Using the boundary conditions $\Phi(0)=\Phi'(0)=0$, we integrate Eq.~\eqref{e:integral3} explicitly:
\begin{equation}
\label{e:integral2}
-\frac{L^2}{2}\left( \Phi'(u)\right)^2=\frac{e^{L^2\Phi(u)}-1}{L\int_0^{\infty} e^{L^2\Phi(v)}dv} .
\end{equation}
Since $\Phi(\infty)=-\infty$ and $\Phi'(\infty)= -1/L$, we conclude that  $L\int_0^{\infty} e^{L^2\Phi(v)}dv=2$ and 
then explicit integration of Eq.~\eqref{e:integral2} gives $\Phi(u)= L^{-2}\cosh^{-2}(Lu/2)$ and 
\be
c(y)=\frac{\P}{2\cosh^2(\P L y/2)}. 
\label{e:c_asympa}
\ee
The corresponding velocity and stress distributions can be also written explicitly.  
If we now combine Eq.~\eqref{e:c_asympa} with Eq.~\eqref{e:velocity_stress_integ_a} we find a simple asymptotic expression 
for the cell velocity $V=\mathcal{P}/2$.
Further substitution of Eq.~\eqref{e:c_asympa} into Eq.~\eqref{e:velocity_stress_integ_b} produces the algebraic equation 
for the cell length~$L$,  
$$
(\P/2) \coth(L/2) =-\mathcal{K} (L-\L) ,
$$
which was quoted in the main text [see Eq.~\eqref{e:collapsed1}]. 


\bibliographystyle{apsrev4-1}


%

\end{document}